\documentclass[reprint, amsmath, amssymb, aps, pra,]{revtex4-2}

\usepackage[pdftex]{graphicx}
\usepackage{dcolumn}
\usepackage{bm}
\usepackage{hyperref}
\usepackage[mathlines]{lineno}
\usepackage{siunitx}

\hypersetup{
    colorlinks=true,
    citecolor=blue,    
    urlcolor=red,
    linkcolor=cyan
}

\begin{document}

\preprint{APS/123-QED}

\title{Dispersion-tuning of nonlinear optical pulse dynamics in gas-filled hollow capillary fibers}

\author{Teodora Grigorova}
\email{t.grigorova@hw.ac.uk}
\author{Christian Brahms}
\author{Federico Belli}
\author{John C. Travers}
\homepage{https://lupo-lab.com/}
\affiliation{School of Engineering and Physical Sciences, Heriot-Watt University, Edinburgh, EH14 4AS, United Kingdom}

\date{\today}

\begin{abstract}
We experimentally investigate the nonlinear optical pulse dynamics of ultrashort laser pulses propagating in gas-filled hollow capillary fibers in different dispersion regimes, which are achieved by tuning the gas pressure. When the pulse propagates in the anomalous dispersion regime we observe soliton dynamics accompanied with soliton-plasma effects, such as self-compression, resonant dispersive-wave emission in the fundamental as well as in higher-order modes, soliton blue-shifting and ionization-induced pulse splitting. Propagation of the pulse in the vicinity of the zero-dispersion wavelength results in pulse splitting and subsequent cross-phase modulation leading to the generation of an additional frequency-shifted band and a 3-octave broad supercontinuum. In the case of pulses propagating in normal dispersion we observe the generation of a broad and flat supercontinuum. In this regime, the experimental results are less well described by simulations that consider only the propagation dynamics inside the fiber. Free-space simulations of the beam propagation in the bulk gas, before the capillary entrance, suggest that this discrepancy is caused by self-focusing and ionization altering the pulse spatial and temporal shape, affecting both the coupling efficiency and the subsequent propagation inside the capillary.
\end{abstract}

\maketitle

\section{Introduction}

Intense ultrashort pulse propagation within a medium is a complex interplay between many linear and nonlinear effects. The group-velocity dispersion (GVD) landscape often determines which of the many nonlinear processes occur, through constraints on phase-matching, and through contributions to the spectral-temporal evolution of the pulse. While the role of GVD has been extensively studied in solid-core optical fibers, especially for supercontinuum generation~\cite{Dudley2006}, in gas-filled photonic-crystal fibers~\cite{Travers2011}, and in bulk material, for filamentation~\cite{Dubietis2017}, it has not been widely studied in gas-filled hollow capillary fibers (HCFs), apart from its role in phase-matching of four-wave mixing \cite{Durfee1997,Misoguti2001}. Gas-filled HCFs have been most widely used for pulse compressors based on spectral broadening through self-phase modulation (SPM) \cite{Nisoli1998}, in experiments which have mostly used short HCF lengths ($\lesssim \SI{1}{\metre}$) of large-core diameter ($\gtrsim \SI{250}{\micro\metre}$), pumped with relatively long pulses ($\gtrsim \SI{30}{\femto\second}$). For these parameters the dispersion in capillaries can be justifiably neglected, since the dispersion of HCFs significantly decreases for larger core sizes, and it does not critically influence the observed dynamics. In contrast, high-order optical soliton dynamics have recently been demonstrated in HCFs~\cite{Travers2019}, primarily by making use of longer HCF lengths, shorter pump pulses, or longer pump wavelengths~\cite{Brahms2020IR}, all of which act to enhance the role of GVD.

In this work, we present an experimental and numerical study of intense ultrafast pulse propagation in HCF over a large gas pressure and pump pulse energy parameter space---corresponding to several fundamentally different dispersion regimes---all within the same optical setup. We achieve this by using short ($\SI{10}{\femto\second}$ duration) pump pulses coupled into a long ($\SI{3}{\metre}$ length), large-core HCF ($\SI{250}{\micro\metre}$ core diameter) filled with argon. Using numerical simulations, which closely reproduce the experimental results in most cases, we study the influence of different physical processes on the characteristic features seen in different spectra and how their contribution is affected by the dispersion landscape in which ultrashort pulses are propagating. In particular, when the pump pulse propagates in the anomalous dispersion regime we observe a breadth of soliton dynamics accompanied with soliton-plasma effects, such as self-compression, resonant dispersive-wave emission in the fundamental as well as in higher-order modes, soliton blue-shifting and ionization-induced pulse splitting. The propagation of the pulse in the vicinity of the zero-dispersion wavelength results in pulse splitting and subsequent cross-phase modulation (XPM) leading to the generation of an additional frequency-shifted band and the generation of a 3-octave broad supercontinuum. In the case of pulses propagating in normal dispersion we observe the generation of a broad and flat supercontinuum, and identify the importance of the free-space dynamics of the pump pulse before coupling into the fiber.

Many (but not all) of the dynamics we study in this paper have been previously observed at much lower energy in small-core micro-structured (photonic-crystal and anti-resonant) hollow-core fibers~\cite{Russell2014}. The key benefit of such fibers is that they guide well even with very small core sizes and so are required when working at low pump energies (the $\SI{}{\micro\joule}$ level). In contrast, large-core gas-filled hollow capillary fibers as studied here enable the use of much higher energy and are completely free of guidance resonances. In fact, when the core size is suitably large, and the fibers are perfectly stretched~\cite{Nagy2008}, HCFs are ideal waveguides, with smooth and exceedingly broad guidance from the X-ray to mid-infrared spectra region. The resonances present in micro-structured hollow-core fibers can complicate (and either enhance beneficially or impede) the observed nonlinear optical pulse dynamics~\cite{Sollapur2017, Tani2018}. Even though these resonances are spectrally localised, they can modify the dispersion profile in which the pump pulse is propagating and lead to phase-matched generation of additional spectral components or even impair soliton self-compression by introducing higher-order dispersion and breaking up the self-compressing pulse. 

\begin{figure}[h]
    \centering
    \includegraphics[width=0.45\textwidth]{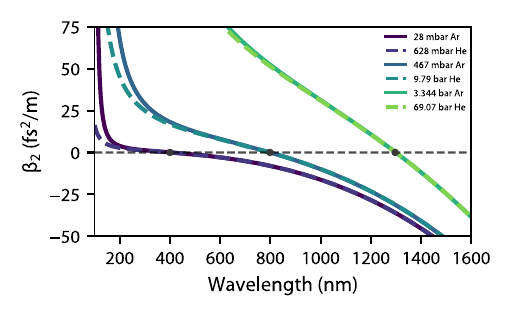}
    \caption{Dispersion tuning in argon- and helium-filled HCF with $\SI{250}{\micro\metre}$ core diameter: changing the argon gas pressure from $\SI{28}{\milli\bar}$ to $\SI{3344}{\milli\bar}$, and helium gas pressure from $\SI{628}{\milli\bar}$ to $\SI{69.07}{\bar}$ tunes the dispersion regime in which a pump pulse at $\SI{800}{\nano\metre}$ is propagating from anomalous to normal. A grey dot represents the zero-dispersion wavelength for the three argon and helium pressures shown: $\SI{400}{\nano\metre}$ for $\SI{28}{\milli\bar}$ Ar or $\SI{628}{\milli\bar}$ He, $\SI{800}{\nano\metre}$ for $\SI{467}{\milli\bar}$ Ar or $\SI{9.79}{\bar}$ He, and $\SI{1300}{\nano\metre}$ for $\SI{3344}{\milli\bar}$ Ar or $\SI{69.07}{\bar}$ He.}
    \label{fig:dispersion_tuning}
\end{figure}

Our dispersion tuning relies on the characteristic GVD profile of gas-filled HCFs as a function of wavelength $\lambda$ and gas pressure $p$, given by~\cite{Marcatili1964, Travers2019}:
\begin{equation}
    \beta_{2}(\lambda, p)  = \frac{\lambda^3}{4 \pi c^2} \left( \rho (p) \frac{\partial^2 \chi_\text{e}}{\partial \lambda^2} - \frac{u_{nm}^2}{2 \pi^2 a^2} \right)
    \label{eq:gvd}
\end{equation}
with $\beta_2$ being the GVD, $\rho$ being the pressure-dependent gas density relative to some standard conditions, $\chi_\text{e}$ being the susceptibility of the filling gas species at those standard conditions available through Sellmeier equations~\cite{Borzsonyi2008, Ermolov2015}, $u_\mathrm{nm}$ the $m^{\text{th}}$ zero of the Bessel function $J_\mathrm{n-1}$, and $a$ the core radius of the HCF. Setting $m=n=1$ selects the fundamental HE$_{11}$ mode. There are two contributions to the dispersion of a gas-filled HCF: the dispersive filling gas and the waveguide dispersion, given by the first and the second term in Eq.~\ref{eq:gvd}, respectively. The waveguide dispersion of an evacuated capillary fiber is anomalous ($\beta_2 < 0$) for all wavelengths, while the gas dispersion is normal ($\beta_2 > 0$) in the ultraviolet to near-infrared range and can be controlled by the gas pressure, shown in Fig.~\ref{fig:dispersion_tuning}. Noble gases (such as helium, neon, argon, krypton, and xenon) have dispersion curves with similar shapes across the visible and near-infrared spectral region, but with different absolute magnitudes. They can be made to approximately match, or be tuned in a similar way, by tuning the gas pressure. Consequently, the results we show in this paper for argon are readily transferable to other gases. Larger differences do appear in the ultraviolet spectral region, closer to the electronic resonances, and do play a role when generating very deep or vacuum ultraviolet light. The dispersion landscape of gas-filled HCFs can be parameterized by the zero-dispersion wavelength $\lambda_\mathrm{zd}$ (ZDW), defined by $\beta_2(\lambda_\mathrm{zd}) = 0$, and the pump wavelength $\lambda_\mathrm{p}$. For the dispersion profile of HCF, this means anomalous dispersion for pump wavelengths longer than the ZDW ($ \lambda_\mathrm{p} > \lambda_\mathrm{zd}$) and normal dispersion for shorter pump wavelengths ($\lambda_\mathrm{p} < \lambda_\mathrm{zd}$). In this work we tune the ZDW from $\SI{400}{\nano\metre}$ to $\SI{1300}{\nano\metre}$ by varying the argon pressure in the HCF in the range from $\SI{28}{\milli\bar}$ to $\SI{3344}{\milli\bar}$. 

This paper is structured as follows. In the next section we describe the experimental and numerical methods that we use. In Section~\ref{sec:anomalous} we describe results for pumping in the anomalous dispersion regime. In Section~\ref{sec:zero} we describe pumping around the zero dispersion wavelength, and in Section~\ref{sec:normal} we describe our results for pumping in the normal dispersion region.

\section{Methods}

\subsection{Experimental setup}

To observe soliton dynamics in capillaries with pump pulses at a central wavelength of around $\SI{800}{\nano\metre}$ requires an experimental arrangement which differs from most other nonlinear optics experiments performed previously in capillaries, in that we require shorter pump pulses or longer fiber lengths. Following Ref.~\cite{Travers2019}, we employ two HCF stages, the first stage solely used to compress the pulses available from our laser system and a second stage, where we observe the dynamics of ultrashort pulse propagation.

A detailed description of the setup can be found in Ref.~\cite{Travers2019}. In summary, we start with $\SI{3}{\milli\joule}$, $\SI{1}{\kilo\hertz}$, $\SI{26}{\femto\second}$-long linearly polarised pulses at a central wavelength of $\SI{800}{\nano\metre}$, produced by a commercial Ti:Sapphire oscillator and amplifier system. As a first stage, we use a conventional HCF compressor, consisting of a $\SI{1.7}{\metre}$-long, $\SI{450}{\micro\metre}$ core diameter stretched HCF, mounted in a gas cell, which is filled with $\SI{2.2}{\bar}$ helium, followed by 6 pairs of double-angle chirped mirrors for phase compensation and a pair of thin silica wedges for fine-tuning. At the output of the compressor the pulses have a nearly transform-limited full-width at half-maximum duration of $\SI{10}{\femto\second}$ as characterised with a home-built second-harmonic-generation frequency-resolved optical gating (FROG) device. The pulse energy can be controlled from a few hundreds of $\SI{}{\nano\joule}$ up to $\SI{1.05}{\milli\joule}$ with a variable attenuator consisting of a broadband $\lambda/2$-waveplate and Brewster reflection from a silicon plate. The compressed pulses are then coupled into a second, $\SI{3}{\metre}$-long, $\SI{250}{\micro\metre}$ core diameter stretched HCF filled with argon, where we vary the pressure from $\SI{28}{\milli\bar}$ to $\SI{3344}{\milli\bar}$, which tunes the ZDW in the range from $\SI{400}{\nano\metre}$ to $\SI{1300}{\nano\metre}$. The theoretical linear throughput of this stage is $\SI{66}{\percent}$ and the coupling efficiency is $\SI{65}{\percent}$, estimated from the measured transmission of the evacuated HCF. However, we estimate from FROG measurements that only around $\SI{80}{\percent}$ of the pulse energy is within the $\SI{10}{\femto\second}$ compressed pulse, effectively decreasing the available energy budget that contributes to the nonlinear dynamics of the pump pulses in the second stage. The remaining part of the pulse energy is not properly compressed, and manifests in a spiky remnant in the pump spectrum, which does not play a role in the observed nonlinear pulse dynamics.

At each pressure we vary the input energy and record the output spectrum and output energy of the second HCF stage with 2 spectrometers (measurement ranges $\SI{200}{\nano\metre}-\SI{1080}{\nano\metre}$ and $\SI{900}{\nano\metre}-\SI{1700}{\nano\metre}$), connected to an integrating sphere and calibrated as a complete system over the entire measurement range (with spectrally calibrated deuterium and tungsten lamps). The spectra from the two spectrometers are combined at $\SI{900}{\nano\metre}$ simply by scaling the two recorded spectra to the used integration times and applying a pre-measured calibration curve. No other post-processing is done; the output of the HCF system is stable over the time necessary to record such an energy scan ($\sim \SI{10}{\minute}$).

\subsection{Numerical simulations}

The experimental data we record is the spectrum at the output of the fiber at a set gas pressure as a function of pump pulse energy. We simulate our experiments by numerically propagating laser pulses over the energy range which we record in the experiment. This allows us to infer the propagation dynamics from the simulations, provided the experimental and the simulated spectra agree closely.

The simulations of pulse propagation through the HCF use our open source full-field multi-mode unidirectional pulse propagation code \cite{Brahms_Luna,Travers2019,Brahms2021}. This model includes modal dispersion and loss, intermodal coupling, full vector medium polarisation model, photoionization (included using PPT ionization rates) and plasma dynamics. The code has been tested and extensively used for pulse propagation simulations with a wide range of parameters previously. We have neglected spatially non-local ionization effects \cite{Koehler2018}, since the pulses we use are sufficiently separated in time, at a repetition rate of \SI{1}{\kilo\hertz}. We do not use any free parameters in our simulations. We use the experimental fiber parameters and filling gas pressure, and the input and coupled energy have been estimated from the measured vacuum HCF transmission. We use $\SI{10}{\femto\second}$ analytical $\text{sech}^2$-shaped pulses as input, based on the pulse duration obtained from FROG measurements. While using the FROG-retrieved pulses as input to the simulations would have given better quantitative agreement between simulation and experiment, we wanted to highlight the general nonlinear optical pulse dynamics, and hence have opted for non-modulated, well-defined analytical input pulses. The difference between simulation and experiment can be observed most clearly around the pump wavelength at $\SI{800}{\nano\metre}$, but overall the dynamics are reproduced well. For these simulations, we initialised the input fully in the fundamental (HE$_{11}$) mode of the fiber, but we modelled the nonlinear propagation dynamics through the fiber for six modes (HE$_{1m}$ with $m$ up to 6). The simulations allow us to separately extract the spectral and temporal fields generated in each of these higher-order modes through modal coupling. The multi-modal spectra or temporal profiles are obtained as a superposition of the modes in the corresponding domain: $S(\omega, z) = \sum_{j} |E_j(\omega, z)|^2$ or $P(t, z) = \sum_{j} |E_j(t, z)|^2$, where $E_j$ is the modal flux in the mode HE$_{1j}$ in either the spectral or temporal domains, $S(\omega)$ is the multi-modal spectral power density, and $P(t)$ is the multi-modal time-domain instantaneous power.

While we routinely achieve good agreement between experimental results and simulations, some experimental conditions have proven harder to emulate numerically. A common assumption when modelling nonlinear pulse propagation in fibers is that the pump pulses do not exhibit any spatial, spectral, or temporal modifications before being coupled into the fiber core, and that all of the measured pump pulse energy is coupled into the fundamental mode of the fiber. However, this assumption breaks down whenever the pump pulse peak power approaches the critical power for self-focusing $P_\mathrm{cr}$ or when the pump intensity leads to significant photoionization. In these cases the free-space propagation of the pump pulse before coupling to the HCF would lead to its modification and should be taken into account. In addition, for the higher gas pressures used in our experiments, the pulse can broaden due to gas dispersion or experience self-phase modulation even before entering the HCF. To capture these effects, in Section \ref{sec:normal}, we additionally model the free-space propagation of the pump pulses before being coupled to the HCF, using a cylindrical-symmetric version of the unidirectional pulse propagation equation, which is also included in the pulse propagation code used \cite{Brahms_Luna}. More details about these simulations can be found in Appendix \ref{ap:appendix}. We simulate the free-space propagation only over the last 20 cm before the fiber entrance, where the beam has reached sufficient intensity for self-focusing and ionization to start to play a role. After the field is propagated in this way to the entrance of the fiber, we have calculated the excitation of HCF modes by using the non-normalised overlap integral \ref{eq:overlap} between the full-spatial electric field at the input of the fiber and of the different fiber modes, which gives the frequency- and time-dependent modal excitation for a chosen number of modes. The field in these modes is then used for multi-modal propagation simulation through the HCF.

\section{Results}

\subsection{Pumping in the anomalous dispersion regime}
\label{sec:anomalous}

\begin{figure*}[t]
    \centering
    \includegraphics[width=\textwidth]{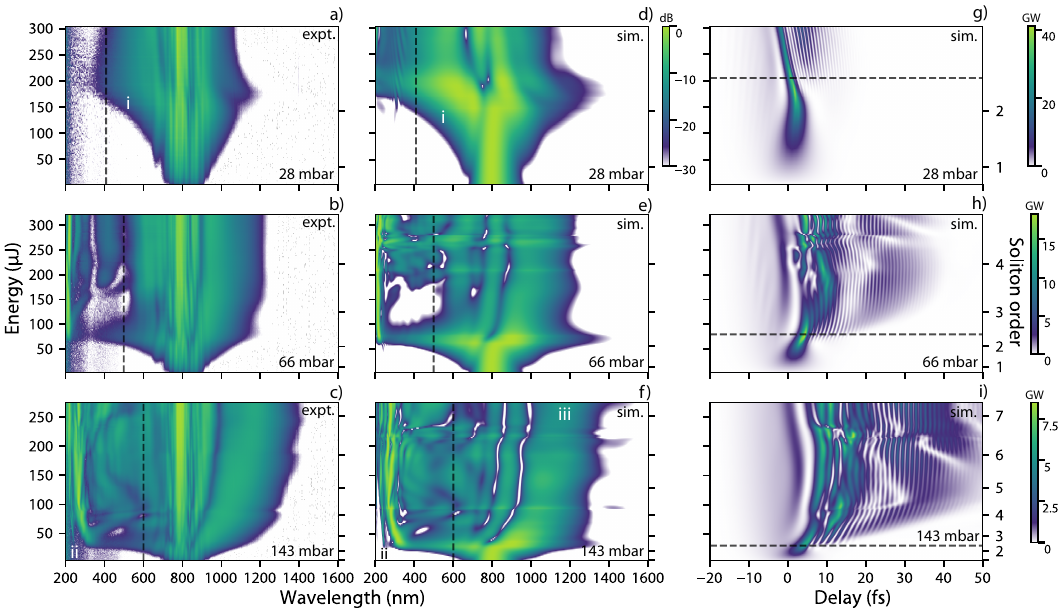}
    \caption{Experimental and simulated energy scans of pump dynamics in the anomalous dispersion regime:  a) -- c) experimental output spectra, d) -- f) simulated multi-mode output spectra, g) -- i) simulated  multi-mode temporal profiles for a range of different pump energies, at 3 different Ar filling pressures: a), d), g) -- $\SI{28}{\milli\bar}$; b), e), h) -- $\SI{66}{\milli\bar}$; and c), f), i) -- $\SI{143}{\milli\bar}$; corresponding to ZDWs (indicated by a vertical dashed grey line in a)--f)) of $\SI{400}{\nano\metre}$, $\SI{500}{\nano\metre}$ and $\SI{600}{\nano\metre}$, respectively. The left y-axis in a)--i) is in terms of pump pulse energy and the right y-axis is the corresponding soliton order $N$. The horizontal dashed gray line in g)--i) indicates $N = 2.4$. Specific features are indicated as i: blue-shifting soliton, ii: HOM-RDW, iii: recoil soliton}
    \label{fig:anomalous_dispersion}
\end{figure*}

First, we look at the case when the pump pulse propagates well into anomalous dispersion regime. As mentioned previously, this corresponds to a pump wavelength longer than the ZDW of the gas-filled fiber; in our case ZDWs shorter than $\SI{800}{\nano\metre}$. The propagation of ultrashort pulses in media with positive nonlinear refractive index $n_2$ and anomalous dispersion results in the formation of temporal optical solitons. An important parameter in this regime is the corresponding soliton order, which is a measure of the balance between the effect of SPM and GVD:
\begin{equation}
    N^2 = \frac{L_\text{d}}{L_\text{nl}},
    \label{eq:soliton_order}
\end{equation}
with $L_\text{d} = \tau_0^2/|\beta_2|$ and $L_\text{nl} = 1/\gamma P_0$, the dispersion and the nonlinear lengths, being the characteristic length scales for GVD and SPM, respectively. $\gamma$ is the nonlinear coefficient (adjusted for gas density) and $\tau_0$ is the natural duration of the pulse, e.g.~$\tau_0 = 0.567 \tau_\text{fwhm}$ for a $\text{sech}^2$-shaped input pulse \cite{Agrawal}. Together with the ZDW, the soliton order can be used as a measure of how much the dynamics are dominated by SPM in both the anomalous and normal dispersion regime \cite{Zheltikov2018}. $N$ is, however, not well defined close to the ZDW, where $L_\text{d}$ approaches infinity.

Propagation in the anomalous dispersion regime is important as a method of generating single- and sub-cycle pulses and the emission of tunable frequency up-conversion. Previously, the observation of soliton-effect self-compression and resonant dispersive wave emission (RDW) in HCF using both $\SI{800}{\nano\metre}$ pump pulses \cite{Travers2019}, and further in the infrared using $\SI{1800}{\nano\metre}$ pump pulses \cite{Brahms2020IR}, even in a very compact setup \cite{Brahms2019mini} has been demonstrated. Further, the generation of circularly polarised RDW was demonstrated \cite{Lekosiotis2021}, as well as the use of pressure gradient HCF setups to allow dispersion-free delivery to in-vacuum targets \cite{Brahms2020gradient}.

Fig.~\ref{fig:anomalous_dispersion} shows a selection of new experimental results and corresponding simulations, which are chosen to represent some useful and well-studied soliton dynamics: self-compression \cite{Travers2011, Mak2013rdw}, soliton-plasma interactions \cite{Hoelzer2011, Chang2013, Koettig2017, Saleh2012} and emission of RDW radiation in the fundamental and higher-order modes (HOM) \cite{Mak2013selfcompression, Tani2014}. While all of the observed effects in this regime have been studied elsewhere in the literature on the platform of hollow-core photonic-crystal fibers or anti-resonant fibers, we want to outline the variety of possible dynamics and to emphasise how small differences in dispersion, via the filling gas pressure, can have very significant impact on the observed dynamics. This also shows that not only self-compression and RDW emission are possible to achieve in HCF, as shown in \cite{Travers2019}, but the full variety of soliton-driven effects.

We have tuned the ZDW from $\SI{400}{\nano\metre}$ to $\SI{600}{\nano\metre}$ by changing the filling argon pressure from $\SI{28}{\milli\bar}$ to $\SI{143}{\milli\bar}$, and scan the energy of the pump pulse coupled in the HCF for each pressure. Fig.~\ref{fig:anomalous_dispersion} a)--c) show our experimental results, and d)--f) show numerical simulations in the spectral domain and g)--i) in the time domain.

The dynamics with $\SI{28}{\milli\bar}$ Ar filling pressure in Fig.~\ref{fig:anomalous_dispersion} a), d) and g) are typical for soliton-plasma interactions in the low soliton order regime---we observe a blue-shifting soliton, ejected from the pump pulse, that shifts from $\SI{800}{\nano\metre}$ to approximately $\SI{600}{\nano\metre}$ in the spectral domain and accelerates (moves to negative delays) in the time domain. The regime of soliton-plasma interaction is of significant practical interest, as previous studies have identified these dynamics as an efficient way to obtain tunable ultrashort pulses in the visible and near-IR spectral regions \cite{Huang2019a, Huang2019b, Huang2019c}.

Another process associated with the dynamics of self-compressed solitons is resonant dispersive-wave emission. This is a phase-matched process, and as such it strongly relies on the dispersion profile in a wide spectral range to determine the wavelength at which the emission will take place. Although our simulations for the $\SI{28}{\milli\bar}$ filling argon pressure show the emission of a RDW below $\SI{200}{\nano\metre}$, this is not covered by our measurement range. With increasing filling gas pressure, up to $\SI{66}{\milli\bar}$, shown in Fig.~\ref{fig:anomalous_dispersion} b), e) and h), the phase-matched wavelength is tuned to longer wavelengths and we can experimentally measure RDW emission at around $\SI{210}{\nano\metre}$. Simulations and other experimental studies have shown that the RDWs are emitted in short pulses \cite{Ermolov2016, Brahms2019duv, Kotsina2022} and are a promising technique to generate tunable ultrafast radiation in the deep and vacuum ultraviolet spectral regions by tuning the filling gas pressure. The RDW emission process is fully coherent and can be highly stable, even in the presence of energy- and duration-jitter with non-CEP-stabilised pump pulses \cite{Brahms2021}. However, since they are emitted at a wavelength in the region of normal dispersion of the gas-filled HCF, they usually quickly broaden, and their duration at the output of the fiber is longer than their transform-limited duration. This can be counteracted by using negative pressure gradients (high pressure at the input of the HCF, vacuum at the output of the HCF), which have been shown to maintain close to transform-limited RDW pulse duration at the fiber output \cite{Brahms2020gradient, Brahms2019duv}.

Further increasing the filling gas pressure to $\SI{143}{\milli\bar}$ in Fig.~\ref{fig:anomalous_dispersion} c), f) and i) shows the emission of RDWs in higher-order modes (HOMs) $\text{HE}_{12}$ below $\SI{250}{\nano\metre}$, which is shorter than the RDW in the fundamental mode ($\SI{325}{\nano\metre}$ to $\SI{250}{\nano\metre}$ depending on the pump energy). The spectral content in each HOM can be extracted from the simulations. The emission of RDWs happens around the point of soliton self-compression, because it relies on the spectrum of the self-compressed soliton extending to the phase-matched wavelength of the RDW to initiate energy transfer \cite{Akhmediev1995}. The generation of RDWs in higher-order modes often occurs in these experiments \cite{Tani2014}. RDW emission in a higher-order mode does not imply or require that the self-compressing soliton that seeds this emission is in a higher-order mode itself, merely that, at the compression point, a RDW in a HOM is phase-matched. It is also not a result of conventional self-focusing or plasma-induced mode coupling. The Kerr-induced nonlinear polarisation always has HOM content because it is the cube of the field. This process is also different from  the emission of a RDW in a HOM directly from a self-compressing pulse propagating in the same HOM, where the modal content of RDW is inherited from the driving pulse and may lead to phase-matching even shorter RDW wavelengths~\cite{Brahms2022HOMRDW}. RDW emission in HOMs naturally occurs at shorter wavelengths than in the fundamental mode because of the stronger anomalous waveguide dispersion of HOMs, which modifies the phase-matching conditions (higher $u_\mathrm{nm}$ in Eq.~(\ref{eq:gvd})).

During RDW emission a significant portion of the pump pulse energy shifts to longer wavelengths, a well-known process described as soliton recoil \cite{Dudley2006, Erkintalo2012}. For example, for the case of a filling gas pressure of $\SI{143}{\milli\bar}$ Ar and an input pulse energy of $\SI{150}{\micro\joule}$, spectral filtering of the simulated pulse in the range $\SI{930}{\nano\metre}$-$\SI{1600}{\nano\metre}$ at the output of the fiber, indicated as iii in Fig.~\ref{fig:anomalous_dispersion}, shows that this part of the spectral output corresponds to a $\SI{13}{\femto\second}$ chirped pulse, centered at around $\SI{1050}{\nano\metre}$, carrying $\SI{31}{\percent}$ ($\SI{18.8}{\micro\joule}$) of the total output energy of $\SI{60.5}{\micro\joule}$.

Another interesting observation is that the temporal dynamics are mostly preserved for the largest range of input energies (from approximately $\SI{150}{\micro\joule}$ to $\SI{200}{\micro\joule}$) for the case of pump pulses propagating in the lower pressure regime, as seen by comparing the temporal dynamics in Fig.~\ref{fig:anomalous_dispersion} g)-i) for three gas pressures shown. This is partly due to the fact that the soliton number in this regime is still low ($N<3$) even for high input energies, because of the reduced nonlinearity and more strongly anomalous dispersion.  

\begin{figure}
    \centering
    \includegraphics[width=\columnwidth]{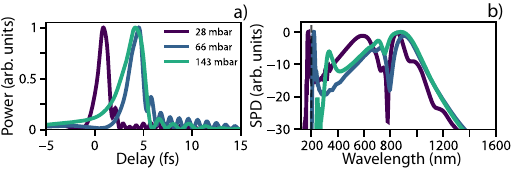}
    \caption{The difference in the self-compression for the same soliton order $N=2.4$ at the output of the HCF, but for different ZDWs: 400 (28 mbar Ar), 500 (66 mbar Ar) and 600 nm (143 mbar Ar), as a) instantaneous power in the time-domain and b) spectral power density (SPD).}
    \label{fig:self_compression}
\end{figure}

One of the most promising applications of systems operating in the soliton regime is their use for pulse compression and the generation of optical transients~\cite{Travers2019}. Optimisation conditions for achieving sub-cycle optical transients are discussed in Ref.~\cite{Voronin2014}. To summarise---to achieve the shortest possible pulse a broad spectral range of anomalous dispersion is required to support soliton propagation without perturbations from higher-order dispersion, along with high peak pump power so that self-steepening and plasma formation enhance the blue side of the spectral broadening. The peak power is capped by the need to avoid high ionization losses, pulse break-up or modulation instability. Optimisation criteria for the soliton order to achieve best self-compression have also been discussed in other works~\cite{Travers2011, Schade2021}, with a primary conclusion being that clean sub-cycle pulses are achieved when pumping with pre-compressed few-cycle pulses and a low soliton order. However, even when keeping the soliton order the same, different compression and different dynamics can be achieved based on different dispersion landscapes, as shown in Fig.~\ref{fig:self_compression}. Fig.~\ref{fig:self_compression} a) shows the temporal intensity profile of the pulse at the output of the HCF for the same soliton order $N=2.4$, albeit different pulse energy, obtained from simulations (as shown with horizontal lines in Fig.~\ref{fig:anomalous_dispersion} g)-i)).  Fig.~\ref{fig:self_compression} b) shows the corresponding spectra. In all three cases the pulse self-compresses to a full-width half-maximum envelope duration below a single cycle at a carrier of $\SI{800}{\nano\metre}$ - $\SI{1.22}{\femto\second}$ for $\SI{28}{\milli\bar}$, $\SI{1.59}{\femto\second}$ for $\SI{66}{\milli\bar}$ and $\SI{2.44}{\femto\second}$ for $\SI{143}{\milli\bar}$ filling gas pressure. We observe the cleanest and shortest self-compressed pulse in the case of $\SI{28}{\milli\bar}$ Ar-filling pressure. This can be associated with the role of the ionization-induced blue-shifting soliton and the fact that the ZDW is further away from the soliton's central wavelength, which creates a wider spectral window where the soliton can self-compress without the perturbation of phase-matched resonant dispersive wave emission. The acceleration caused by the plasma blue-shift is also clearly visible in the earlier arrival time of the pulse in the case of $\SI{28}{\milli\bar}$ Ar. The low nonlinearity also helps to keep the soliton order low---enabling clean self-compression---even at high energies.

\subsection{Pumping near or at the zero-dispersion wavelength}
\label{sec:zero}

\begin{figure*}[t]
    \centering
    \includegraphics[width=\textwidth]{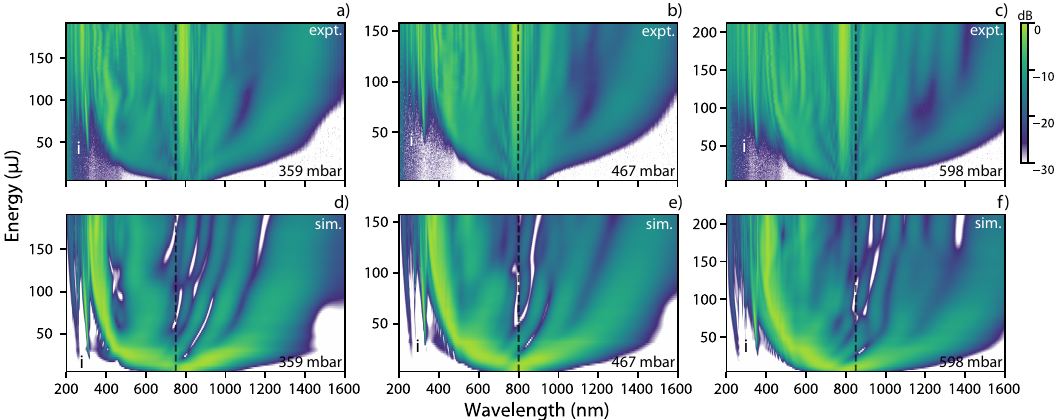}
    \caption{Experimental and simulated energy scans of the pump dynamics close to zero dispersion: a)-c) experimental output spectra, d)-f) stimulated multi-mode output spectra for a range of different Ar filling pressures: a), d) - $\SI{359}{\milli\bar}$; b), e) - $\SI{467}{\milli\bar}$; and c), f) - $\SI{598}{\milli\bar}$; corresponding to ZDWs (indicated by a vertical dashed grey line in a)-f)) of $\SI{750}{\nano\metre}$, $\SI{800}{\nano\metre}$ and $\SI{850}{\nano\metre}$, respectively. Radiation in the HOMs is indicated by i.}
    \label{fig:nearZDW}
\end{figure*}

Quite different propagation dynamics occur when the pump pulse is propagating near or at the ZDW of the gas-filled HCF. This range is also used for purely-SPM spectral broadening in HCF-based compressors, where it is usually assumed that the dispersion does not have a major influence on the dynamics of the propagating pulse. However, when using shorter pulses, which have a broader bandwidth (e.g. a $\SI{10}{\femto\second}$ pulse has a transform-limited bandwidth of approximately $\SI{100}{\nano\metre}$), the variation of the dispersion around the ZDW does matter. Some of our experimental results and simulations are shown in Fig.~\ref{fig:nearZDW}. The filling argon pressure was varied between $\SI{359}{\milli\bar}$ (Fig.~\ref{fig:nearZDW} a) and d)) and $\SI{598}{\milli\bar}$ to tune the ZDW from $\SI{750}{\nano\metre}$ (Fig.~\ref{fig:nearZDW} c) and f)) to $\SI{850}{\nano\metre}$. A ZDW of $\SI{800}{\nano\metre}$ is obtained for an Ar filling pressure of $\SI{467}{\milli\bar}$ (Fig.~\ref{fig:nearZDW} b) and e)). It can be seen that the experimentally measured dynamics as a function of pump energy are very similar in all three cases. Three-octave spanning supercontinua, from $\SI{200}{\nano\metre}$ to more than $\SI{1600}{\nano\metre}$, can be achieved in this regime and the shift of a significant part of the spectrum to longer wavelengths is notable. The major part of the observed dynamics is happening in the fundamental mode, even though in this parameter range a lot of capillary modes are excited and our numerical simulations show us that the narrow-band radiation below $\SI{400}{\nano\metre}$ is emitted in HOMs (marked i in Fig.~\ref{fig:nearZDW}). However, there is an increased discrepancy between the simulations and the experimental measurements, even though the overall trends are recovered. Particularly, simulations show an increased transfer of energy to wavelengths around $\SI{400}{\nano\metre}$ in Fig.~\ref{fig:nearZDW} d)-f).

\begin{figure*}[t]
    \centering
    \includegraphics[width=\textwidth]{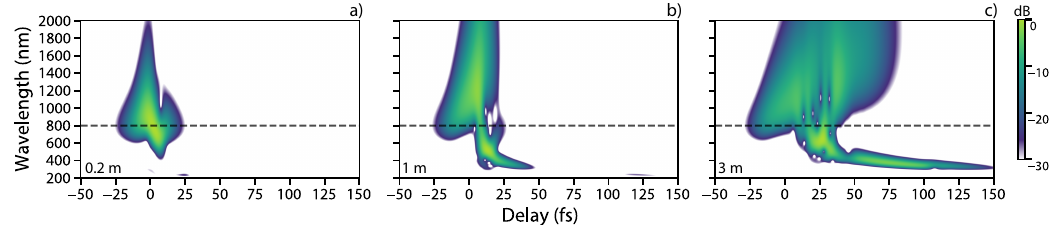}
    \caption{Spectograms of the propagation dynamics of a $\SI{160}{\micro\joule}$ pulse in $\SI{467}{\milli\bar}$ Ar-filled HCF (ZDW $\SI{800}{\nano\metre}$) in the fundamental HE$_{11}$ mode calculated using a $\SI{4}{\femto\second}$ gate pulse a) after $\SI{0.2}{\metre}$ of propagation along the fiber, b) after $\SI{1}{\metre}$ of propagation, and c) after $\SI{3}{\metre}$ of propagation at the output of the HCF. The horizontal line indicates the ZDW, which is $\SI{800}{\nano\metre}$.}
    \label{fig:spectrograms}
\end{figure*}

Insight into the dynamics of the generation of this spectral band can be obtained from spectrograms, which show the connection between the temporal and spectral domain \cite{Dudley2006}. Spectrograms at different points during the propagation along the HCF for $\SI{160}{\micro\joule}$ pump energy and $\SI{467}{\milli\bar}$ Ar pressure are shown in Fig.~\ref{fig:spectrograms} for three different positions: $\SI{0.2}{\metre}$ in a), $\SI{1}{\metre}$ in b), and $\SI{3}{\metre}$ in c). This representation allows us to extract the sequence of dynamics that lead to the emission of the high-frequency radiation. It can be seen from Fig.~\ref{fig:spectrograms} a) that initially the pulse broadens due to SPM, which leads to two parts of the pulse propagating in different dispersion regimes---normal for wavelengths shorter than the ZDW at $\SI{800}{\nano\metre}$ and anomalous for longer wavelengths---while their group velocities remain nearly matched, similarly to previous studies of pulse propagation near the ZDW of a fiber~\cite{Wai1987}. Further along the pulse propagation, after $\SI{1}{\metre}$, shown in Fig.~\ref{fig:spectrograms} b), the two parts of the pump pulse, propagating in opposite dispersion regimes, interact through XPM and the short-wavelength edge at approximately $\SI{400}{\nano\metre}$ is enhanced. At the output of the fiber, at $\SI{3}{\metre}$ in Fig.~\ref{fig:spectrograms} c), the band of newly generated short-wavelength radiation is temporally well separated from the rest of the pump, since it has lower group velocity and is delayed with respect to the rest of the pulse.

This sequence of dynamics has been discussed in Ref.~\cite{Skryabin2005}, suggesting that the part of the pulse that remains in the anomalous dispersion region forms a soliton (fundamental or higher-order). However, we have not observed that this part of the pulse exhibits soliton properties, so formation of a soliton in the anomalous dispersion regime is not necessary for the second stage of the propagation dynamics, in which the now separated pulses interact with each other through XPM. In recent years, XPM has been considered analogously to an event horizon effect~\cite{Philbin2008}, temporal reflection~\cite{Plansinis2018}, as well as front-induced transition~\cite{Gaafar2019}. Temporal reflection of a probe pulse off a soliton has previously also been suggested as a mechanism for the generation of octave-spanning supercontinua~\cite{Demircan2013} and similar propagation dynamics have recently been observed experimentally in solid-core fibers in the mid-IR region~\cite{Ghosh2019}.

\begin{figure}[t]
    \centering
    \includegraphics[width=0.5\textwidth]{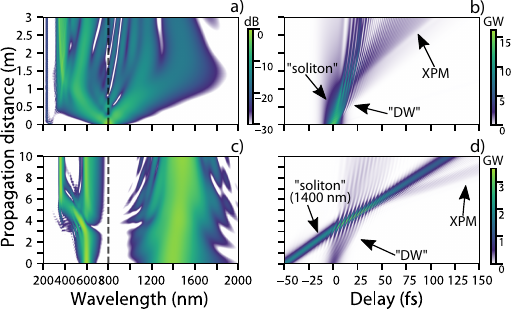}
    \caption{Comparison of the propagation dynamics with simplified model: a), b) the simulated spectral and temporal evolution in HE$_{11}$ mode during propagation of a $\SI{10}{\femto\second}$, $\SI{800}{\nano\meter}$, $\SI{160}{\micro\joule}$ pulse in a  $\SI{125}{\micro\metre}$ core radius HCF and length of $\SI{3}{\metre}$, filled with $\SI{467}{\milli\bar}$ Ar (corresponding to the experimental geometry). c), d) the simulated spectral and temporal dynamics of two pulse collision (pulse 1, ``DW" (dispersive wave): $\SI{10}{\femto\second}$ FWHM duration, central wavelength $\SI{600}{\nano\metre}$, $\SI{5}{\micro\joule}$  pulse energy; pulse 2, ``soliton": $\SI{10}{\femto\second}$ FWHM duration, central wavelength $\SI{1400}{\nano\metre}$, $\SI{20}{\micro\joule}$; pulse 2 is initially delayed with respect to pulse 1 by $\SI{50}{\femto\second}$) in a HCF of the same core radius, and same filling gas pressure, but a length of $\SI{10}{\metre}$ (the loss of the fiber has been ignored in this case). ``XPM": pulse reflected due to XPM.}
    \label{fig:toy_model}
\end{figure}

To clarify this explanation, in Fig.~\ref{fig:toy_model} we compare the propagation dynamics of the previously discussed case to a simplified scenario involving the interaction and temporal collision of two pulses inserted with initial conditions chosen to highlight the dynamics. Fig.~\ref{fig:toy_model} a) and b) show the spectral and temporal dynamics corresponding to our experiment, with a single $\SI{10}{\femto\second}$, $\SI{160}{\micro\joule}$, pump pulse at a central wavelength of $\SI{800}{\nano\meter}$, in an HCF filled with $\SI{467}{\milli\bar}$ Ar. Fig.~\ref{fig:toy_model} c) and d) shows the spectral and temporal dynamics of our simplified scenario, consisting of two $\SI{10}{\femto\second}$ duration pulses: the probe pulse (also referred to as a ``dispersive wave" in the associated literature~\cite{Skryabin2005}, but distinct from RDWs) with central wavelength of $\SI{600}{\nano\metre}$ and $\SI{5}{\micro\joule}$ pulse energy, and the pump pulse (also referred to as the ``soliton") with central wavelength of $\SI{1400}{\nano\metre}$ and $\SI{20}{\micro\joule}$ pulse energy. The second pulse is initially delayed with respect to the first by $\SI{50}{\femto\second}$ in order to clearly observe the interaction between the two pulses when they temporally overlap. They are propagated in an HCF of the same core radius and same filling gas pressure as Fig.~\ref{fig:toy_model} a) and b), but with a much longer length of $\SI{10}{\metre}$ (while neglecting loss) to make the collision dynamics clearly observable despite the small difference in group velocity between the two pulses. In the first (experimental) case, the initial separation between these pulses is much smaller and hence they collide after a shorter distance.

In this simplified scenario the single original input pulse has been separated into two pulses which propagate in two different dispersion regimes---the ``soliton" (or pump) pulse corresponds to the part of the input pulse that is shifted to the anomalous dispersion regime (longer than the ZDW) after the initial SPM step, and the ``dispersive wave" (or probe) pulse corresponds to the part of the input pulse that shifts into the normal dispersion region. The pulse-edge on which the XPM interaction happens is very important (leading or trailing edge of the pulses), because it determines whether the XPM pulse will be blue- or red-shifted. We see that the blue-shifted part of the spectrum of the probe pulse appears after the two pulses collide in the time domain. This blue-sifted part appears as reflected from the pump pulse in the temporal domain. Although there are some significant differences between this simplified scenario and the actual dynamics, the major qualitative features are captured, and it supports the proposed sequence of evolution of an ultrashort pulse propagation near the ZDW of an HCF. We have additionally confirmed that the emission of this blue-shifted radiation band does not occur in simulations in the absence of interaction between these two pulses, i.e if any of the two pulses propagate on their own inside the fiber.

\subsection{Pumping in the normal dispersion region}
\label{sec:normal}

\begin{figure*}[t]
    \centering
    \includegraphics[width=\textwidth]{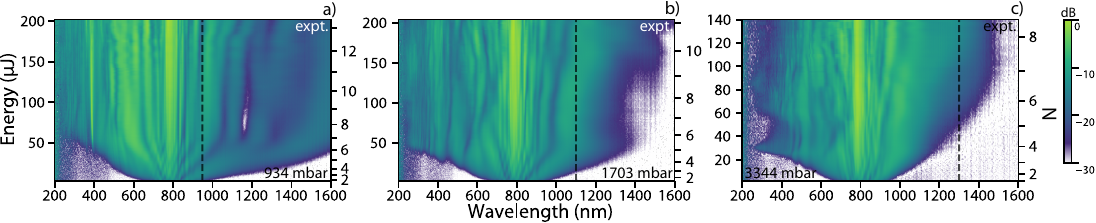}
    \caption{Experimental energy scans of the pump dynamics in normal dispersion regime. Output spectrum as a function of coupled pulse energy for: a) $\SI{934}{\milli\bar}$ Ar, ZDW $\SI{950}{\nano\metre}$; b) $\SI{1703}{\milli\bar}$ Ar, ZDW $\SI{1100}{\nano\metre}$; c) $\SI{3344}{\milli\bar}$ Ar, ZDW $\SI{1300}{\nano\metre}$. The corresponding ZDWs are indicated by a vertical dashed grey line in a)-c). The left y-axis in a)-c) is in terms of pump pulse energy and the right y-axis is the corresponding ``soliton order" $N$.}
    \label{fig:normal_dispersion}
\end{figure*} 

Lastly, with a further increase of the gas filling pressure, the ZDW shifts into the infrared region and the pump pulse propagates in a normal dispersion regime. Our experimental results for $\SI{934}{\milli\bar}$, $\SI{1703}{\milli\bar}$ and $\SI{3344}{\milli\bar}$ filling Ar pressures are shown in Fig.~\ref{fig:normal_dispersion}. They correspond to ZDWs in the range from $\SI{950}{\nano\metre}$ to $\SI{1300}{\nano\metre}$. They show significant spectral broadening, which starts to decrease with increasing gas pressure. However, in this regime our usual simulations, which use the pump pulse fully coupled to the fundamental mode of the HCF, do not qualitatively represent the measurements well, which suggests that some of the assumptions we have used are no longer justified.

One of our assumptions is that all of the energy that is initially coupled into the fiber is in the fundamental mode $\mathrm{HE}_{11}$, despite the fact that hollow capillaries are inherently multi-mode. This is justified for low nonlinearity if the correct focusing geometry is used~\cite{Harrington1998}, where over 98\% of a Gaussian pump beam can be coupled to the fundamental mode. Furthermore, the numerical model we use fully accounts for the nonlinear multi-mode dynamics during the pulse propagation through the fiber. However, in the higher-pressure regime it is possible that the pulse experiences self-focusing even before it is coupled into the fiber, and we have not accounted for this up to now. Indeed, we exceed the peak power for self-focusing~\cite{Fibich2000} for some of the pressures and pump pulse energies used. In the case of ultrashort pulses propagating in free space with normal dispersion, the dynamics are complicated further by the addition of temporal effects~\cite{Polynkin2013, Chernev1992}.

In order to capture these dynamics, we have additionally performed full spatial simulations of the pulse propagation inside the input gas cell over a distance of $\SI{20}{\centi\metre}$ from the fiber input. Although the distance between the gas-cell window and the fiber input is $\SI{92}{\centi\metre}$ in our setup, we have found that significant nonlinear modifications in the spatial and temporal domains occur only near the fiber input, where the intensity of the focusing beam is sufficiently high. As input for these simulations, we have used an aberration-free Gaussian beam with beam size back-calculated from the linear fiber input beam radius of $\SI{80}{\micro\metre}$ corresponding to 0.64 times the inner capillary radius. The free-space propagated beam is then projected onto the modes of the HCF using Eq.~\ref{eq:overlap}, which couples the spatio-temporal evolution due to the free-space propagation onto up to 20 azimuthally-symmetric $\mathrm{HE}_\mathrm{1m}$ fiber modes.

\begin{figure}[t]
    \centering
    \includegraphics[width=0.5\textwidth]{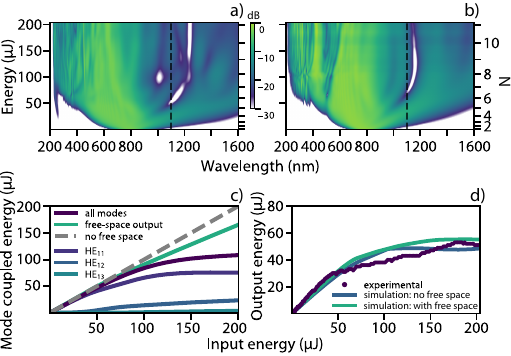}
    \caption{Simulation with and without free-space propagation (with ionization included) before the input of the fiber and modal decomposition for $\SI{1703}{\milli\bar}$ filling Ar pressure: a) without free-space propagation, all of the input energy coupled in the fundamental mode, b) with free-space propagation and overlap of the free-space-propagated beam with fiber modes. c) Different input energies used in the simulations including free-space propagation (output of the free-space propagation shown with label ``free-space output" and the total energy coupled into the fiber after the free-space propagation shown as ``all modes") and not including free-space propagation (shown with ``no free space" label) as a function of the initial pulse energy for the simulation. The energies coupled in the first 3 most excited modes---from $\text{HE}_{11}$ to $\text{HE}_{13}$---at the input of the fiber are also shown. The decrease of the input energy in the case of simulations including free-space propagation is due to ionization losses. d) Output energy from both the simulations with and without free-space propagation and the experimentally measured output energy.}
    \label{fig:modal_decomposition}
\end{figure}

A comparison of fiber propagation simulations with and without the free-space propagation for $\SI{1703}{\milli\bar}$ Ar pressure are shown in Fig.~\ref{fig:modal_decomposition}. Fig.~\ref{fig:modal_decomposition} a) shows a simulated fiber-propagation energy scan for the same parameters as the experimental energy scan shown in Fig.~\ref{fig:normal_dispersion} b). In comparison, Fig.~\ref{fig:modal_decomposition} b) shows the case where the pulse has first been propagated in free space  to the fiber input and then modified spatio-temporal beam profile has been used to couple to the fiber modes. Fig.~\ref{fig:modal_decomposition} c) shows how much energy has been coupled to each mode and Fig.~\ref{fig:modal_decomposition} d) shows the output energy from both of the simulations and of the experimental energy scan.

While neither of the simulations agree very well with experiment, this is not unexpected in this regime with complex spatio-temporal coupling. The biggest discrepancy between the simulations and experimental results is in the infrared region, where the simulations significantly overestimate the spectral broadening. Some parameter adjustment might be needed to get the right correspondence between experiment and simulation, but in this study we chose to use our best estimates of the experimental parameters: the vacuum measured coupling and the measured pulse duration. Another possible reason for the discrepancy is that the laser beam used in our experiments is also not a perfect aberration-free Gaussian beam, which will have an effect on the details of self-focusing and modal coupling. The simulations are radially symmetric, whereas any asymmetry in the actual beam profile can be enhanced by the self-focusing and cause similar asymmetry in the fiber coupling. A further possibility is the formation of long-lived gas-density variations due to plasma recombination and subsequent heating~\cite{Jhajj2014, Koehler2021}, a process not included in our simulations. However, further investigation of the influence of these effects is outside the scope of this work. 

\begin{figure*}[t]
    \centering
    \includegraphics[width=\textwidth]{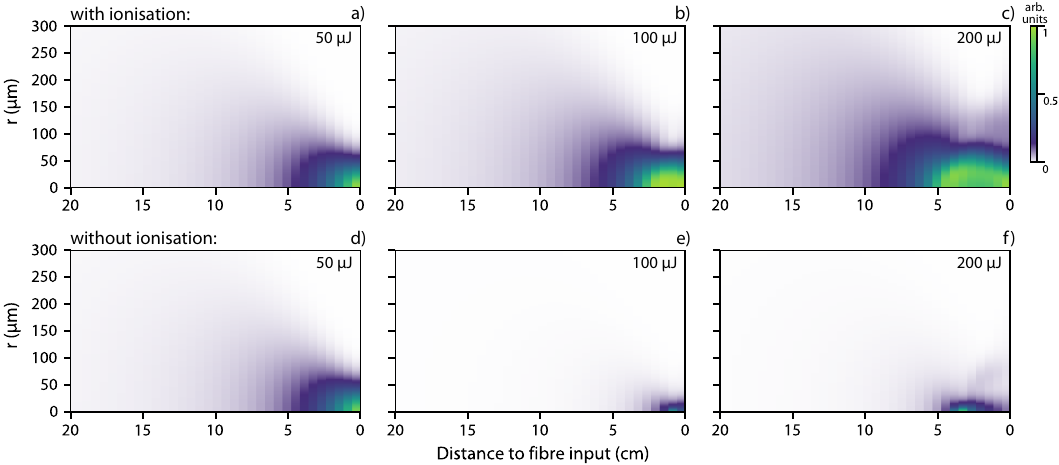}
    \caption{Simulation of the free-space pulse propagation dynamics in $\SI{1703}{\milli\bar}$ Ar for different input energies: a) and d) $\SI{50}{\micro\joule}$, b) and e) $\SI{100}{\micro\joule}$, and c) and f) $\SI{200}{\micro\joule}$. The intensity of the pulse is shown as a function of the distance from the HCF input on a linear scale, normalised to the maximum intensity during the propagation range. The simulations in a)-c) include the effect of ionization during the pulse propagation, while d)-f) do not include ionization. The fiber input is at 0 cm.}
    \label{fig:freespace_prop}
\end{figure*}

From the free-space simulations for the pulse propagation in $\SI{1703}{\milli\bar}$ Ar
we find that there is significant spatial and temporal modification of the pump pulse even before it reaches the input of the HCF. This is shown in Fig.~\ref{fig:freespace_prop} which presents a simulation of the radial beam profile integrated over all frequencies as the beam focuses towards the HCF input for three pump pulse energies---$\SI{50}{\micro\joule}$, $\SI{100}{\micro\joule}$, and $\SI{200}{\micro\joule}$. We consider two cases---with and without including the effect of ionization in the free-space propagation---in order to get a better picture of the effects that play a role in this stage of the dynamics. In both cases, we observe that increasing the energy of the pump pulse leads to the beam focus moving before the fiber input, and a divergent beam at the HCF input. The effect of self-focusing is more distinctly observed in the case which does not include ionization. However, ionization acts to arrest the self-focusing and results in stabilisation of the focal spot size. Although ionization has a distinct qualitative effect on the dynamics, it does not lead to significant losses for the parameter range used here, as seen in Fig.~\ref{fig:modal_decomposition} c). For the maximum pump pulse energy used, $\SI{200}{\micro\joule}$, the critical power for self-focusing $P_\mathrm{cr}$~\cite{Fibich2000} is exceeded by 2.6 times; the pump pulse peak power is equal to $P_\mathrm{cr}$ for an energy $\approx \SI{77}{\micro\joule}$. This suggests that intensity clamping is not the reason for the observed throughput saturation in Fig.~\ref{fig:modal_decomposition} d). The spatial modification of the beam due to self-focusing and ionization causes progressively higher coupling into HOMs as the pump pulse energy is increased, shown in Fig.~\ref{fig:modal_decomposition} c). However, most of the energy coupled in the fiber is still in the fundamental and $\mathrm{HE}_{12}$ mode, and coupling to further HOMs never exceeds $\SI{5}{\micro\joule}$ at the input of the fiber.

\begin{figure}[h]
    \centering
    \includegraphics[width=0.5\textwidth]{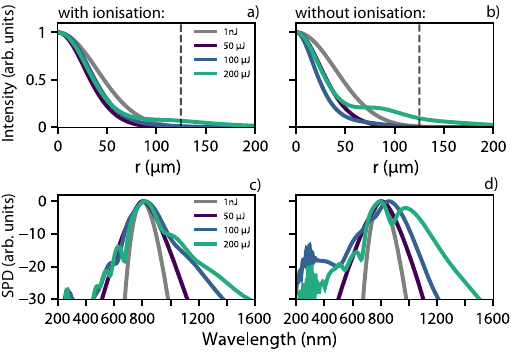}
    \caption{Simulation of the  a), b) spatial and c), d) on-axis spectral modification of the HCF input beam due to free-space propagation in $\SI{1703}{\milli\bar}$ Ar for different input energies---$\SI{50}{\micro\joule}$, $\SI{100}{\micro\joule}$, and $\SI{200}{\micro\joule}$; a) and c) include the effect of ionization during the propagation, while b) and d) do not include ionization. The linear beam focusing is shown in grey, calculated similarly to the other cases but for pulse energy of $\SI{1}{\nano\joule}$, well below the onset of nonlinear effects. The vertical dashed grey line in a) and b) marks the core radius of the HCF used.}
    \label{fig:freespace_fiber_input}
\end{figure}

A summary of the effect of free-space propagation on the beam spatial profile and on-axis spectrum is shown in Fig.~\ref{fig:freespace_fiber_input}. In the spatial domain the focused beam size is reduced due to self-focusing. This decrease is less significant in the simulations including ionization effects, which should represent the true propagation dynamics more closely. In both the simulations with and without including the effect of ionization, at the high end of the used pump pulse energy range, the beam develops a pedestal structure, extending beyond the boundaries of the HCF core, due to divergence after the back-shifted focus. In addition to ionization losses, the drop in coupled energy in the simulations including free-space as well as fiber propagation is due to a combination of spatial and spectral-temporal dynamics, which are difficult to completely disentangle. Lastly, Fig.~\ref{fig:freespace_fiber_input} c) and d) shows the additional spectral broadening of the pulse at the input of the HCF, which results purely from the free-space propagation of the beam. 

While the dynamics in certain regimes have complicated spatio-temporal coupling, they are still of interest for applications and research. Multi-modal propagation has continuously attracted scientific interest, with studies of multi-mode solitonic structures in the anomalous dispersion regime~\cite{Lopez-Zubieta2018a, Lopez-Zubieta2018b}, multi-dimensional molecular interactions~\cite{Safaei2020}, inter-modal four-wave-mixing~\cite{Piccoli2021}, self-collapse and multi-mode self-compression~\cite{Crego2019}, among others. We have shown that in certain situations the free-space propagation of the pump pulse before the entrance to the fiber also plays a role in the dynamics that we observe experimentally, and the fundamental mode coupling assumption generally used is not applicable. For simulations this necessitates additionally including free-space propagation calculations and using a multi-modal input for the HCF propagation. Additionally, all of the presented dynamics are scalable to lower or higher energies, using the scaling laws derived in Ref.~\cite{Travers2019}.

In this work we have studied the optical propagation dynamics in statically-filled HCF, where the gas pressure is constant along the fibre and in the free-space regions between the fibre and the gas-cell windows. An increasing gas-pressure gradient through the HCF can alternatively be used to decrease the ionisation and self-focusing before the input of the fibre, as used in many works on pulse-compression based on SPM in HCF~\cite{Suda2005, Bohman2008, Bohle2014, Nagy2020, Ouille2020}. Similarly, a decreasing gas-pressure gradient can be used for delivery of self-compressed pulses directly to vacuum~\cite{Brahms2020gradient}. In the current work we concentrated on a statically-filled HCF as it is the simplest system to explore the dynamics. Furthermore, our numerical simulations provide some insight into understanding the deterioration in the coupling due to propagation before the fibre, which has not previously been studied in detail.

The dynamics we have observed in the normal dispersion regime were stable in the input pulse energy range considered, while leading to the generation of a multi-octave supercontinuum. For long-wavelength ZDW ($\lambda_\mathrm{zd} \gtrsim \SI{1200}{\nano\metre}$) and input energies higher than the ones shown in Fig.~\ref{fig:normal_dispersion} c), we observe the onset of instabilities in the output spectrum.

\section{Conclusion}

\begin{figure}[h]
    \centering
    \includegraphics[width=0.5\textwidth]{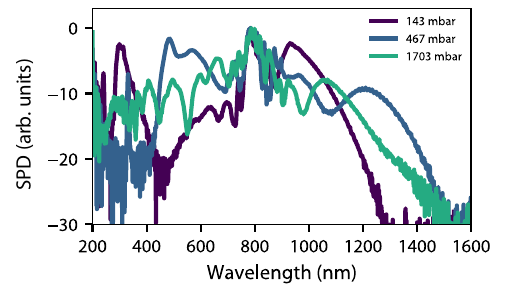}
    \caption{Experimental spectra recorded for pump pulse propagation in different dispersion regimes: anomalous dispersion for $\SI{143}{\milli\bar}$, near-zero dispersion for $\SI{467}{\milli\bar}$, and normal dispersion for $\SI{1703}{\milli\bar}$ filling argon pressure, all for input energy of $\SI{50}{\micro\joule}$.}
    \label{fig:spec_lineouts}
\end{figure}

We have experimentally and numerically studied the dynamics of ultrashort pulses propagating in a gas-filled hollow capillary fiber for a wide range of dispersion landscapes, characterised by the dispersion experienced by the pump, and we have explored the continuous transition between different regimes. The different dispersion at the pump gives rise to qualitatively different regimes of supercontinuum generation, as shown in Fig.~\ref{fig:spec_lineouts}, which are interesting and useful for different applications. In summary, when ultrashort pulses are propagating in the anomalous dispersion regime we observe a breadth of processes related to the dynamics of perturbed higher-order soliton propagation, such as self-compression, ionization-induced blue-shift, and resonant dispersive-wave emission in the fundamental and in higher-order HCF modes. When propagating near the zero-dispersion point of the gas-filled HCF, an ultrashort pulse experiences pulse-splitting and consequently cross-phase modulation between the two pulses, which can lead to the generation of a new band of radiation. Lastly, when propagating in a normal dispersion regime, the ultrashort pulse can still lead to the generation of multi-octave supercontinua, however in this case the free-space propagation of the focusing pulse might additionally have to be considered, leading to more complicated spatio-temporal dynamics, both before and during the propagation in the hollow capillary fiber.  

\begin{acknowledgments}

This work was funded by the European Research Council (ERC) under the European Union’s Horizon 2020 research and innovation programme: Starting Grant agreement HISOL no. 679649 and ERC Consolidator Grant XSOL no. 101001534. This work used EPCCs Cirrus HPC Service (\href{https://www.epcc.ed.ac.uk/cirrus}{https://www.epcc.ed.ac.uk/cirrus}). C.B. acknowledges support from the Royal Academy of Engineering through Research Fellowship No. RF/202122/21/133. F.B. acknowledges support from the Royal Academy of Engineering through Research Fellowship No. RF/202021/20/310.

\end{acknowledgments}

\appendix*
\label{ap:appendix}

\section{Free-space propagation simulations}
Nonlinear propagation in free space is treated in the same framework and by the same code as propagation in the HCF. The model for guided-wave propagation, including the form of the nonlinear source terms and references for material properties, is described in detail in Refs.~\cite{Brahms_Luna, Travers2019, Brahms2021}. For free-space propagation, projection onto the modes of the waveguide is replaced with a transverse $0^\mathrm{th}$-order Hankel transform and propagation losses are neglected, so that the unidirectional pulse propagation equation reads
\begin{multline}
    \partial_z E(\omega, k_\perp, z) = i\left(k_z(\omega, k_\perp) - \frac{\omega}{v}\right)E(\omega, k_\perp, z) \\ + \frac{i \mu_0 \omega^2}{2 k_z(\omega, k_\perp)} P^\mathrm{nl}(\omega, k_\perp, z)\,,
\end{multline}
where $k_z(\omega, k_\perp) = \sqrt{k_0^2 - k_\perp^2} = \sqrt{\frac{\omega^2}{c^2}n^2(\omega) - k_x^2 - k_y^2}$ with $n(\omega)$ the refractive index of the gas, and
\begin{equation}
    E(\omega, k_\perp, z) = 2\pi\int_{-\infty}^{+\infty}\mathrm{d} t \int_0^\infty r \mathrm{d} r E(t, r, z) J_0(k_\perp r) \mathrm{e}^{i\omega t}\,,
\end{equation}
where $J_0(x)$ denotes the $0^\mathrm{th}$-order Bessel function of the first kind. A similar equation applies to the nonlinear polarisation $P^\mathrm{nl}(\omega, k_\perp, z)$. The Hankel transform is implemented using the quasi-discrete Hankel transform~\cite{Yu1998} with a pupil radius of $\SI{10}{\milli\metre}$.

We run this simulation for the final \SI{20}{\cm} of propagation before the HCF entrance, as the beam size before this point is too large and the intensity too low to cause significant nonlinear effects. As the initial condition we back-propagate an aberration-free Gaussian beam with $1/\mathrm{e}^2$ intensity radius of \SI{80}{\um}, which corresponds to 0.64 times the HCF core radius \cite{Harrington1998}, from the HCF entrance to the beginning of the propagation window.

After propagating to the plane of the HCF entrance $z_\mathrm{HCF}$, we determine the field coupled into each mode $j$ of the HCF by calculating the overlap integral with the azimuthally symmetric HE$_{1m}$ modes up to $m=20$:
\begin{equation}
    E_j(\omega, 0) = 2\pi \int^\infty_0 \boldsymbol{E}(\omega, r, z_\mathrm{HCF}) \cdot \boldsymbol{\hat{e}}^*_j(r)\, r \mathrm{d}r\,,
    \label{eq:overlap}
\end{equation}
where $\boldsymbol{\hat{e}}_j(r)$ is the normalised modal field distribution of mode $j$.

\bibliography{refs}

\end{document}